\documentclass[prd,aps,showpacs,notitlepage,superscriptaddress,nofootinbib]{revtex4-1}

\usepackage{amsmath}
\usepackage{multirow}
\usepackage{float}
\usepackage{soul}
\usepackage{caption}
\usepackage {tikz}
\usepackage{gnuplot-lua-tikz}

\def\bea#1\eea{\begin{align}#1\end{align}}

\makeatletter
\def\slash#1{{\mathpalette\c@ncel{#1}}} 
\makeatother

\newcommand\beq{\begin{eqnarray}}
\newcommand\eeq{\end{eqnarray}}

\newcommand\la{\langle}
\newcommand\ra{\rangle}

\begin{document}
\title{The twist-3 gluon contribution to $A_N$ in $J/\psi$ production in $pp$ collisions}

\date{\today}

\author{Longjie Chen}
\email{chenlongjieusc@163.com}
\affiliation{Institute of Nuclear Physics Polish Academy of Sciences,
ul. Radzikowskiego 152, PL-31-342 Krak\'ow, Poland}

\author{Shinsuke Yoshida}
\email{shinyoshida85@gmail.com}
\affiliation{State Key Laboratory of Nuclear Physics and Technology, Institute of
Quantum Matter, South China Normal University, Guangzhou 510006, China}
\affiliation{Guangdong Basic Research Center of Excellence for
Structure and Fundamental Interactions of Matter, Guangdong
Provincial Key Laboratory of Nuclear Science, Guangzhou
510006, China}

\begin{abstract}

We present our results for the twist-3 gluon contribution to
the single transverse-spin asymmetry(SSA) in $J/\psi$ production
in proton-proton collisions. Although the data were reported by the RHIC experiment more 
than a decade ago, a theoretical calculation based on rigorous collinear factorization 
has remained unavailable for the dominant gluonic contribution.
Our results show that only the $C$-even type twist-3 
gluon distribution, which has a direct relationship with 
the gluon TMD distribution function,
contributes to the $J/\psi$ SSA. Therefore, this observable serves as a key probe to understand 
the three-dimensional motion of gluons inside the proton.
We also perform numerical simulations of the $J/\psi$ SSA at RHIC and LHC energies.
Our simulations indicate that a sizable SSA could be generated through a mechanism different from 
that responsible for the SSA in light hadron and $D$-meson production.

\end{abstract}

\maketitle


\section{Introduction}

The understanding of the nucleon's internal structure has made dramatic progress over 
the past half century through constant interactions between theory and experiment. 
At present, those efforts continue to be directed toward 
the next-generation electron-ion collider(EIC) experiment\cite{Accardi:2012qut}.
The central aim of this experiment is to reveal the role of gluons in various aspects
of the nucleon as its main constituents.
Gluons show their unique characteristics in several fundamental problems,
the origin of the mass(trace anomaly), 
the nucleon spin puzzle(gluon spin and the less-known orbital angular momentum) 
and the small-$x$ structure(dense gluonic system). 
One of the significant advantages of the EIC experiment
over previous electron-proton collision experiments is the availability 
of polarized beams for both the initial electron and proton in high energy regimes.
Comprehensive studies of various polarized processes will 
provide a new perspective on our understanding of the nucleon spin structure.
The measurement of the gluon spin inside the proton is one of the major achievements made at 
Relativistic Heavy Ion Collider(RHIC)\cite{PHENIX:2008swq,deFlorian:2014yva}. 
The EIC will focus on the further understanding of the origin of the
proton spin, including the gluon spin contribution from the small-$x$ regimes
and the orbital motion of the gluons.
The single transverse-spin asymmetry has drawn attention recently as a key probe to
investigate such orbital motion of the partons inside the proton.
When large asymmetries were first observed 
in the late 1970s\cite{Klem:1976ui,Bunce:1976yb}, they were regarded as mysteries
in high energy hadron physics since the conventional parton model calculation
could not reproduce them at all\cite{Kane:1978nd}.
The transverse-momentum-dependent(TMD) factorization,
which additionally takes into account the transverse momenta of partons in a scattering process,
has been well developed over the past couple of decades as a possible solution to the large SSAs.
Reasonable descriptions of the existing SSA data within the TMD framework
have made us recognize the importance of the orbital motion of the partons
in high energy hadron reactions. 
It can now be said that SSA has transformed from merely a mystery in QCD
into an ideal tool to investigate the orbital motion of the partons.
Another successful framework for describing the large SSAs
is the twist-3 mechanism in collinear factorization.
This framework describes incoherent multi-parton scatterings
in a hard process and can be applied to single-scale processes such as $pp\to \pi X$ 
at the RHIC. The two frameworks mentioned above have different applicable regions
with respect to the transverse momentum of the particle produced in the final state.
Although they seemingly describe different mechanisms as the origin of the SSA, 
it was found that there exists an intermediate region where the two frameworks 
give consistent cross section formulas for the SSAs in certain processes
\cite{Ji:2006br,Ji:2006vf,Koike:2007dg,Bacchetta:2008xw,Yuan:2009dw,Zhou:2009jm,Ikarashi:2025vdn}. 
This equivalence supports the idea that the two frameworks describe a common origin 
of the large SSAs.

Heavy flavored hadron production serves as an important tool to investigate the gluon structure in the proton
because the heavy quark that fragments into the final state heavy hadron is mainly produced
through gluon fusion process\cite{Qiu:2020xum}.
The orbital motion of gluons can be accessed through measurements of the SSAs
in heavy meson production.
The SSAs in heavy meson production have been extensively studied in recent years within
the TMD framework for both $D$-meson 
production\cite{Anselmino:2004nk,Godbole:2016tvq,DAlesio:2017rzj,Godbole:2017fab,DAlesio:2018rnv} 
and $J/\psi$ production\cite{Godbole:2014tha,Mukherjee:2016qxa,DAlesio:2017rzj,Rajesh:2018qks,DAlesio:2018rnv,Sun:2019tuk,DAlesio:2019qpk,Kishore:2019fzb,DAlesio:2019gnu,DAlesio:2020eqo,Kato:2024vzt}.
On the other hand, the twist-3 gluon contribution to the SSA
was calculated for $D$-meson 
production about 15 years ago\cite{Kang:2008qh,Kang:2008ih,Beppu:2010qn,Koike:2011mb,Beppu:2012vi}
and the first calculation for the $J/\psi$ SSA in semi-inclusive deep inelastic scattering
was done very recently\cite{Chen:2023hvu}. 
The calculation of the SSA in $pp^{\uparrow}\to J/\psi\,X$ has been done only for quark 
contribution \cite{Schafer:2013wca}
although the corresponding data were reported in 2010 by the RHIC 
experiment\cite{PHENIX:2010hqq}. In this paper, we present the first result for the twist-3 gluon contribution 
to the SSA in this process. We also perform numerical simulations of the SSA
at RHIC and LHC energies with a simple model for the twist-3 gluon distribution functions
by taking into account the data points reported in \cite{PHENIX:2010hqq}. 

The remainder of this paper is organized as follows: In section II, we introduce the
definitions of the twist-3 gluon distribution functions relevant to our study and show some required 
relations for the calculation. In section III, we show our result 
of the unpolarized and the polarized cross section formulas within the collinear factorization. 
In section IV, we show numerical simulations for the $J/\psi$ SSA
at RHIC  and LHC energies. Section V is devoted to a summary of our study.
All the materials required for the twist-3 calculation are summarized in the Appendix.


\section{Definitions of twist-3 gluon distribution functions}

We introduce the definitions of the twist-3 gluon distribution functions.
In general, the twist-3 cross section can be expressed in terms of the intrinsic, kinematical and dynamical
twist-3 functions. The contribution from the intrinsic function is canceled in the case of the Sivers effect
because of $PT$-invariance and, therefore, only the other two types are relevant to our study.
The kinematical functions for a transversely polarized proton are defined as follows
\cite{Koike:2019zxc}:
\beq
&&\Phi^{(\pm)\alpha\beta\gamma}_{\partial}(x)
\nonumber\\
&=&i\int{d\lambda\over 2\pi}\,e^{i\lambda x}
\la pS|F^{\beta n}(0)[0,\lambda n]D^{\gamma}(\lambda n)F^{\alpha n}(\lambda n)|pS\ra 
\nonumber\\
&&-\int{d\lambda\over 2\pi}\,e^{i\lambda x}
\int^{\mp\infty}_{\lambda}d\mu\,
\la pS|F^{\beta n}(0)[0,\mu n]gF^{\gamma n}(\mu n)[\mu n,\lambda n]F^{\alpha n}(\lambda n)|pS\ra 
\nonumber\\
&=&{M_N\over 2}g^{\alpha\beta}_{\perp}\epsilon^{pnS_{\perp}\gamma}G^{(\pm)(1)}_T(x)
+i{M_N\over 2}\epsilon^{pn \alpha\beta}S_{\perp}^{\gamma}\Delta G^{(\pm)(1)}_T(x)
+{M_N\over 8}\Bigl(\epsilon^{pnS_{\perp}\{\alpha}g^{\beta\}\gamma}_{\perp}
+\epsilon^{pn\gamma\{\alpha}S^{\beta\}}_{\perp}\Bigr)\Delta H^{(\pm)(1)}_{T}(x)+\cdots,\hspace{2mm}
\label{kinematical}
\eeq
where the index $\gamma$ is restricted to the transverse components and
$p, S_{\perp}$ and $M_N$ respectively denote the proton's momentum, spin and mass. 
We use the simplified notation 
$\epsilon^{pnS_{\perp}\gamma}=\epsilon^{\mu\nu\rho\gamma}p_{\mu}n_{\nu}S_{\perp\rho}$. 
$[0,\lambda n]$ is a shorthand notation of the gauge-link operator in the adjoint representation,
\beq
[0,\lambda n]\equiv {\rm P}\exp\Bigl(ig\int_{\lambda}^0dt\, A^n(tn)\Big),
\eeq
which guarantees the gauge-invariance of the matrix elements.
The vector $n$ is light-like and satisfies $p\cdot n=1, n^2=0$.
Naively $T$-odd observables such as the SSA receive contributions from 
$G^{(1)}_T(x)$ and $\Delta H^{(1)}_{T}(x)$.

The dynamical gluon distribution functions are defined by matrix elements composed
of three gluon field strength tensors\cite{Ji:1992eu,Beppu:2010qn}.
The dynamical functions are categorized into two types, $C$-even function $N(x_1,x_2)$ and 
$C$-odd function $O(x_1,x_2)$, reflecting the fact that there are two structure constants $if^{abc}$
and $d^{abc}$ in SU($N_c$) group,
\beq
N^{\alpha\beta\gamma}(x_1,x_2)&=&i\int{d\lambda\over 2\pi}\int{d\mu\over 2\pi}
e^{i\lambda x_1}e^{i\mu (x_2-x_1)}
\la pS_{\perp}|if^{bca}F_{b}^{\beta n}(0)
gF_c^{\gamma n}(\mu n)F_a^{\alpha n}(\lambda n)|pS_{\perp}\ra
\nonumber\\
&=&2iM_N\Bigl[g_{\perp}^{\alpha\beta}\epsilon^{\gamma pnS_{\perp}}N(x_1,x_2)
-g_{\perp}^{\beta\gamma}\epsilon^{\alpha pnS_{\perp}}N(x_2,x_2-x_1)
-g_{\perp}^{\alpha\gamma}\epsilon^{\beta pnS_{\perp}}N(x_1,x_1-x_2)\Bigr]+\cdots,
\label{C-even}
\eeq
\beq
O^{\alpha\beta\gamma}(x_1,x_2)&=&i\int{d\lambda\over 2\pi}\int{d\mu\over 2\pi}
e^{i\lambda x_1}e^{i\mu (x_2-x_1)}
\la pS_{\perp}|d^{bca}F_{b}^{\beta n}(0)
gF_c^{\gamma n}(\mu n)F_a^{\alpha n}(\lambda n)|pS_{\perp}\ra
\nonumber\\
&=&2iM_N\Bigl[g_{\perp}^{\alpha\beta}\epsilon^{\gamma pnS_{\perp}}O(x_1,x_2)
+g_{\perp}^{\beta\gamma}\epsilon^{\alpha pnS_{\perp}}O(x_2,x_2-x_1)
+g_{\perp}^{\alpha\gamma}\epsilon^{\beta pnS_{\perp}}O(x_1,x_1-x_2)\Bigr]+\cdots,
\label{C-odd}
\eeq
where we omitted gauge-links for simplicity. These dynamical functions satisfy 
the following symmetries.
\beq
O(x_1,x_2)&=&O(x_2,x_1),\hspace{5mm}O(x_1,x_2)=O(-x_1,-x_2),
\nonumber\\
N(x_1,x_2)&=&N(x_2,x_1),\hspace{5mm}N(x_1,x_2)=-N(-x_1,-x_2).
\label{symmetries}
\eeq
The following relations were derived in \cite{Koike:2019zxc}. 
\beq
G_T^{(\pm)(1)}(x)=\pm 4\pi(N(x,x)-N(x,0)),\hspace{5mm}\Delta H_T^{(\pm)(1)}(x)=\mp 8\pi N(x,0).
\label{relations}
\eeq
Using these relations, the polarized cross section can be written solely in terms of the dynamical functions
as shown in the next section.


\section{Calculation of the SSA in $J/\psi$ production in $pp$ collisions}

\subsection{NRQCD framework and the unpolarized cross section for $J/\psi$ production}

We consider $J/\psi$ production in proton-proton collisions,
\beq
p^{\uparrow}(p,S_{\perp})+p(p')\to J/\psi(P_{J})+X.
\eeq
We adopt the nonrelativistic QCD(NRQCD) framework\cite{Caswell:1985ui,Bodwin:1994jh}
for the hadronization of the charm quark pair
into $J/\psi$. The above scattering is illustrated within this framework as
\beq
p^{\uparrow}(p,S_{\perp})+p(p')\to \sum_nc\bar{c}[n](P_J)+X,
\eeq
where $n$ denotes all possible Fock states of the charm quark pair. As a first step, we 
consider the color-singlet contribution $n={}^3S_1^{[1]}$ in which the hadronization 
of $J/\psi$ gives the following structure in the diagram calculation.
\beq
{\cal N}\langle O^{J/\psi}[{}^3S_1^{(1)}]\rangle
[\slash{\epsilon}(\slash{P}_{J}+m_J)]_{ij}\delta_{i'j'},
\eeq
where $\epsilon^{\rho}$ is the polarization vector of $J/\psi$,
$i,j(i',j')$ are the spinor(color) indices, ${\cal N}$
is the normalization factor and $\langle O^{J/\psi}[{}^3S_1^{(1)}]\rangle$
is the long-distance-matrix-element(LDME) which describes the nonperturbative 
transition from the color-singlet charm quark pair into $J/\psi$. 
\begin{figure}[H]
\begin{center}
  \includegraphics[height=4cm,width=3cm]{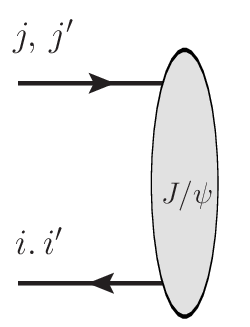}
\end{center}
 \caption{Diagrammatic expression of the hadronization process in the color-singlet contribution.}
\label{color_singlet}
\end{figure}
\noindent
The unpolarized $J/\psi$ gives
\beq
\sum\epsilon^{\rho}\epsilon^{\sigma}=-g^{\rho\sigma}
+{P_{J}^{\rho}P_{J}^{\sigma}\over m_J^2}.
\eeq
\begin{figure}[H]
\begin{center}
  \includegraphics[height=3cm,width=12cm]{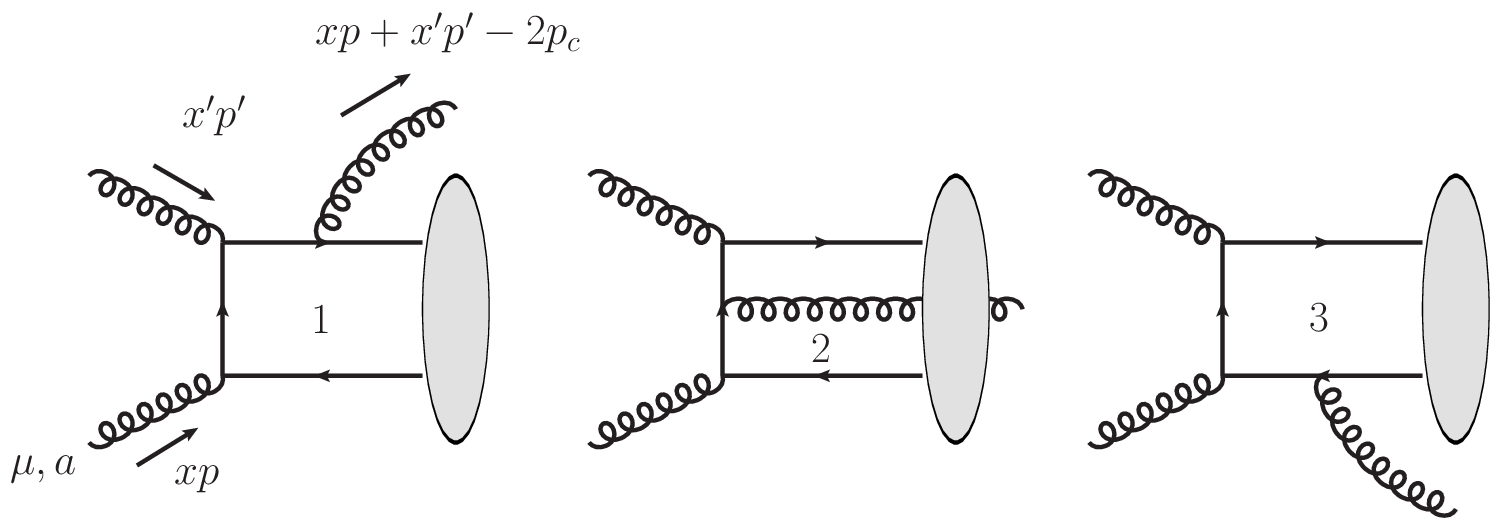}
  \includegraphics[height=3cm,width=9cm]{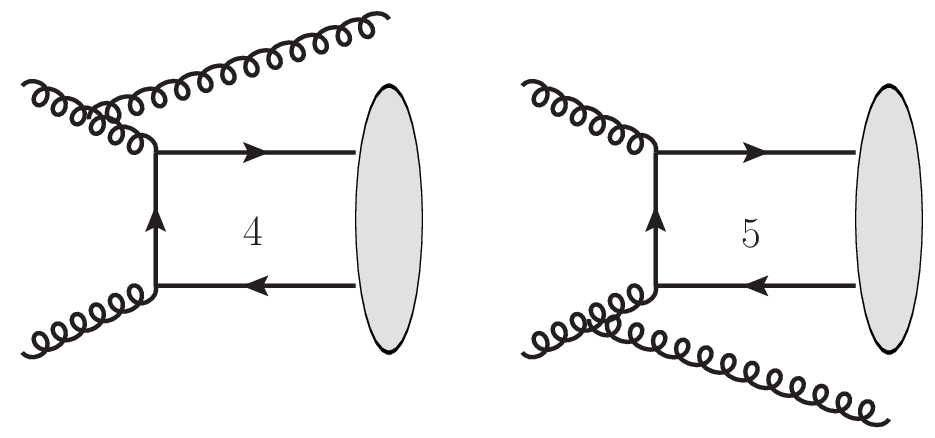}
  \includegraphics[height=3cm,width=12cm]{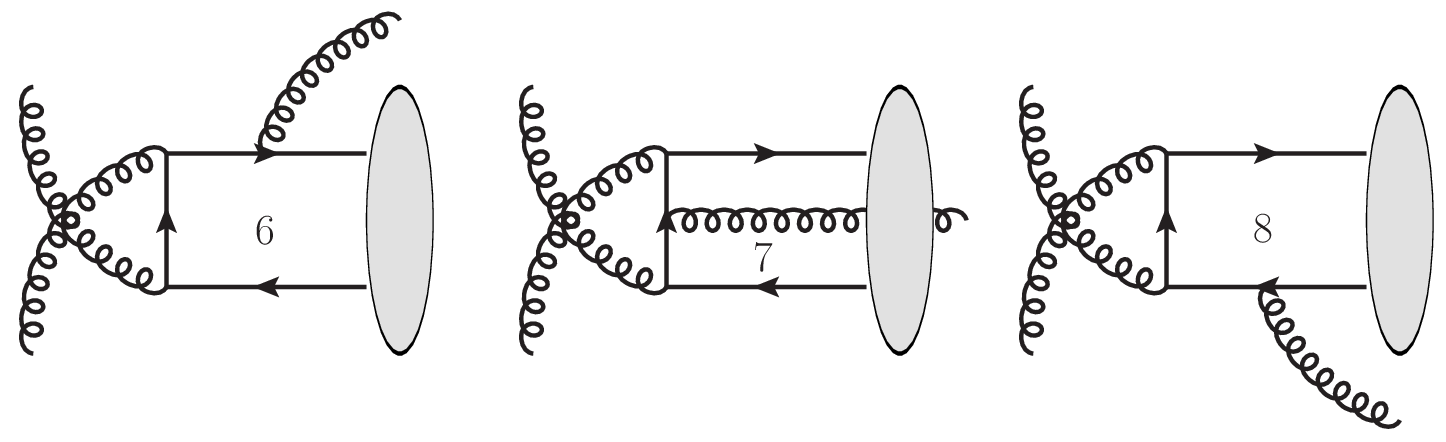}
  \includegraphics[height=3cm,width=9cm]{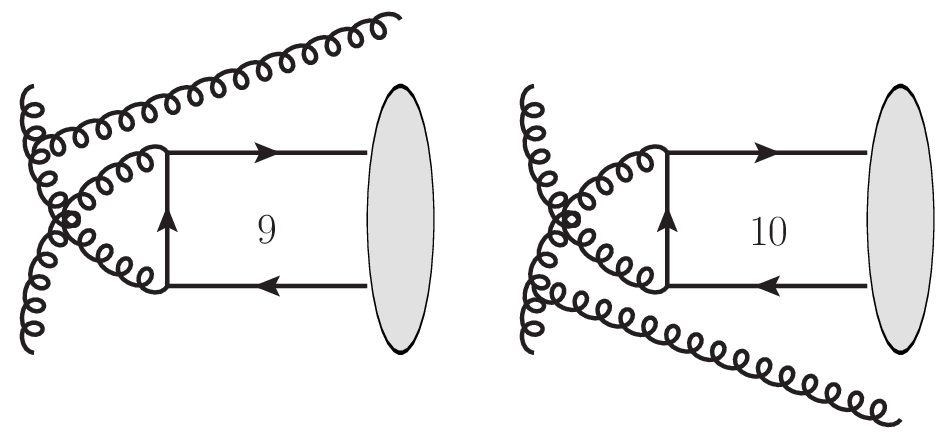}
  \includegraphics[height=3cm,width=9cm]{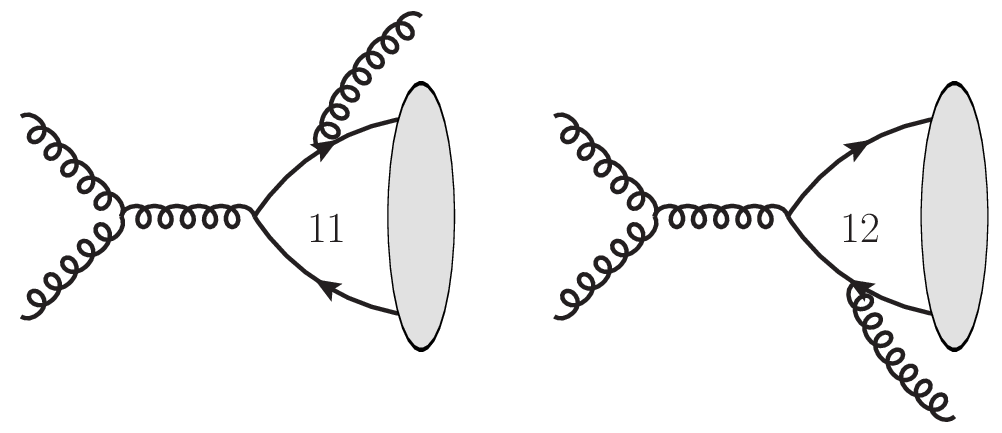}
\end{center}
 \caption{LO diagrams with ${}^3S_1^{(1)}$ charm quark pair in the final state.
 We express each numbered diagram as $C_{i}^a{\cal M}_{i\,\mu}(xp)\ (i=1,2,\cdots, 12)$.}
\label{LO_diagrams}
\end{figure}
\noindent
There are 12 nonvanishing diagrams shown in Fig. \ref{LO_diagrams} at the level of the amplitude.
The unpolarized cross section can be calculated by taking the square of the amplitude as
\beq
P_{J}^0{d\sigma\over d^3P_{J}}
&=&{4\pi\alpha_s^3\over S}\Bigl({\cal N}\langle O^{J/\psi}[{}^3S_1^{(1)}]\rangle\Bigr)
\int{dx\over x}G(x)\int{dx'\over x'}G(x')
\hat{\sigma}^U\delta(\hat{s}+\hat{t}+\hat{u}-m_J^2)
\nonumber\\
&=&{4\pi\alpha_s^3\over S}\Bigl({\cal N}\langle O^{J/\psi}[{}^3S_1^{(1)}]\rangle\Bigr)
{G(x)\over x}\int^1_{x'_{\rm min}}{dx'\over x'}G(x')
{1\over x'S+T-m_J^2}\hat{\sigma}^U\Bigr|_{x=\bar{x}},
\eeq
where $\alpha_s$ is the QCD coupling, $S$ is the center-of-mass energy,
$G(x)$ is the unpolarized gluon distribution function of the proton, $\bar{x}$ is given by
\beq
\bar{x}=-{x'U+m_J^2(1-x')\over x'S+T-m_J^2},\hspace{5mm}
x'_{\rm min}=-{T\over S+U-m_J^2},
\eeq
and Mandelstam variables are defined by
\begin{align}
\begin{aligned}
S &= (p+p')^2, \\
T &= (p-P_J)^2, \\
U &= (p'-P_J)^2,
\end{aligned}
\qquad
\begin{aligned}
\hat{s} &= (xp+x'p')^2 = xx'S, \\
\hat{t} &= (xp-P_J)^2 = xT + m_J^2(1-x), \\
\hat{u} &= (x'p'-P_J)^2 = x'U + m_J^2(1-x').
\end{aligned}
\end{align}
The $i$-th diagram in Fig. \ref{LO_diagrams} can be decomposed into the color part $C^a_{i}$ 
and the tensor part ${\cal M}_{i\mu}$.
The color factor of the squared amplitude is obtained as
\beq
C^a_{i}C^{b*}_{j}=C_{ij}\delta^{ab},\qquad(i,j=1,2,\cdots 12).
\eeq
Thus one can obtain the known result for the hard cross section $\hat{\sigma}^U$ \cite{Baier:1981uk} as
\beq
\hat{\sigma}^U&=&\sum_{i,j=1}^{12}C_{ij}
\Bigl(-{1\over 2}g^{\mu\nu}_{\perp}(p)\Bigr){\cal M}_{i\,\mu}(xp){\cal M}_{i\,\nu}^{*}(xp)
\nonumber\\
&=&\Bigl(-{1\over N^2_cC_F}+{1\over 4C_F}\Bigr)
{128m_J^2\Bigl((\hat{t}^2+\hat{t}\hat{u}+\hat{u}^2)^2
+m_J^4(\hat{t}^2+\hat{t}\hat{u}+\hat{u}^2)-m_J^2(2\hat{t}^3+3\hat{t}^2\hat{u}
+3\hat{t}\hat{u}^2+2\hat{u}^3)
\Bigr)
\over (\hat{s}-m_J^2)^2(\hat{t}-m_J^2)^2(\hat{u}-m_J^2)^2},
\eeq
where $g_{\perp}^{\mu\nu}(p)=g^{\mu\nu}-p^{\mu}n^{\nu}-p^{\nu}n^{\mu}$.

\subsection{Twist-3 polarized cross section for $J/\psi$ production in $pp$ collisions}

We calculate the twist-3 gluon distribution contribution to the polarized cross section.
The basic formulation for deriving the cross section formula was established 
based on the modern nonpole technique in \cite{Yoshida:2022vnf} for $ep$ collisions. 
Some extension is needed in $pp$ collisions because of the fact that there are both 
the initial state interaction(ISI) and the final state interaction(FSI).
We present here the polarized cross section formula for $pp$ collisions and 
the derivation in detail will be discussed in a separate paper because it is quite lengthy and technical.
When a process involves both ISI and FSI, we can write down the formula for the 3-gluon distribution 
contribution to the polarized cross section as
\beq
P_{J}^0{d\Delta\sigma\over d^3P_{J}}
&=&{1\over 16\pi^2 S}\Bigl({\cal N}\langle O^{J/\psi}[{}^3S_1^{(1)}]\rangle\Bigr)
\omega^{\mu}_{\ \alpha}\omega^{\nu}_{\ \beta}\omega^{\lambda}_{\ \gamma}\int{dx'\over x'}G(x')
\nonumber\\
&&\times\Biggl[\int{dx\over x^2}\sum_{i,j=1}^{12}\Bigl(\Phi^{(-)\alpha\beta\gamma}_{\partial}(x)
(C^{f(+)[-]}_{Lij}+C^{f(-)[-]}_{Lij})+\Phi^{(+)\alpha\beta\gamma}_{\partial}(x)C^{f(-)[+]}_{Lij}
\nonumber\\
&&+2\pi C^{d(-)[+]}_{Lij}O^{\alpha\beta\gamma}(x,x)
\Bigl){\partial\over \partial k^{\lambda}}[{\cal M}_{i\mu}(k){\cal M}^{*}_{j\nu}(k)]\Bigr|_{k=xp}
\nonumber\\
&&-{1\over 2}\int dx_1\int dx_2 \Bigl[{-if^{abc}\over N_c(N_c^2-1)}N^{\alpha\beta\gamma}(x_1,x_2)
+{N_cd^{abc}\over (N_c^2-4)(N_c^2-1)}O^{\alpha\beta\gamma}(x_1,x_2)\Bigr]
\nonumber\\
&&\times \Bigl({1\over x_2}{1\over x_1\pm i\epsilon}
{1\over x_2-x_1\mp i\epsilon}{\cal M}^{(\pm)[\mp]ac}_{L\mu\lambda}(x_1p,(x_2-x_1)p)
[\sum_{i=1}^{12}C_{i}^{b*}{\cal M}^{*}_{i\,\nu}(x_2p)]
\nonumber\\
&&+{1\over x_1}{1\over x_2\pm i\epsilon}
{1\over x_2-x_1\mp i\epsilon}[\sum_{i=1}^{12}C^a_i{\cal M}_{i\,\mu}(x_1p)]
{\cal M}^{(\pm)[\mp]bc}_{R\nu\lambda}(x_2p,(x_2-x_1)p)
\Bigr)\Biggr],
\label{polarized}
\eeq
where the color factors $C^{f}_{Lij}$ and $C^{d}_{Lij}$ are shown in the Appendix.
${\cal M}^{(\pm)[\mp]ac}_{L\mu\lambda}(x_1p,(x_2-x_1)p)$
is the sum of the diagrams including an additional gluon line with momentum $(x_2-x_1)p$ 
and there are 86 diagrams as shown in Fig. \ref{dynamical_dots}.
${\cal M}^{(\pm)[\mp]bc}_{R\nu\lambda}(x_2p,(x_2-x_1)p)$ is its complex conjugate.
The complicated pole structure composed of $1/(x_1\pm i\epsilon)$, $1/(x_2\pm i\epsilon)$
and $1/(x_2-x_1\pm i\epsilon)$ reflects the presence of ISI and FSI and 
all contributions have to be combined.
We show the explicit forms of ${\cal M}^{(\pm)[\mp]ac}_{L\mu\lambda}(x_1p,(x_2-x_1)p)$ 
in the Appendix.
After evaluating all diagrams, the polarized cross section is given by
\beq
&&P_{J}^0{d\Delta\sigma\over d^3P_{J}}
\nonumber\\
&=&{4\pi\alpha_s^3(-\pi M_N)\over S}
\Bigl({\cal N}\langle O^{J/\psi}[{}^3S_1^{(1)}]\rangle\Bigr)
\int{dx\over x'}G(x')\int{dx\over x^2}\delta(\hat{s}+\hat{t}+\hat{u}-m_J^2)
\nonumber\\
&&\times\Bigl(\hat{s}\epsilon^{P_JpnS_{\perp}}-(\hat{t}-m_J^2)
x'\epsilon^{npp'S_{\perp}}\Bigr)\Bigl(-{1\over N^2_cC_F}+{1\over 4C_F}\Bigr)
\Bigl[x{d\over dx}N(x,x)\Bigl(-{4\over \hat{s}\hat{u}}\hat{\sigma}^U\Bigr)
+x{d\over dx}N(x,0)\hat{\sigma}_D
\nonumber\\
&&+N(x,x)\hat{\sigma}_{ND1}+N(x,0)\hat{\sigma}_{ND2}
+N(x,Hx)\hat{\sigma}_{H1}+N(x-Hx,x)\hat{\sigma}_{H2}+N(Hx,Hx-x)\hat{\sigma}_{H3}\Bigr],
\label{polarized}
\eeq
where $H=-(\hat{u}-m_J^2)/(\hat{s}-\hat{u})$. We find that 
the contribution from the $C$-odd function $O(x_1,x_2)$ is completely canceled
as in the case of SIDIS\cite{Koike:2011mb}. 
Note that the result is independent of the arbitrary vector $n$,
\beq
\hat{s}\epsilon^{P_JpnS_{\perp}}-(\hat{t}-m_J^2)
x'\epsilon^{npp'S_{\perp}}=2xx'\epsilon^{P_Jpp'S_{\perp}},
\eeq
which follows from the identity
\beq
g^{\alpha\beta}\epsilon^{\mu\nu\rho\sigma}
=g^{\alpha\mu}\epsilon^{\beta\nu\rho\sigma}+g^{\alpha\nu}\epsilon^{\mu\beta\rho\sigma}
+g^{\alpha\rho}\epsilon^{\mu\nu\beta\sigma}+g^{\alpha\sigma}\epsilon^{\mu\nu\rho\beta}.
\eeq
The hard cross sections are given by
\beq
\hat{\sigma}_D&=&{1\over \hat{s}\hat{u}}
{512m_J^2\Bigl((\hat{t}^2+\hat{t}\hat{u}+\hat{u}^2)^2
+m_J^4(\hat{t}^2+\hat{u}^2)-2m_J^2(\hat{t}^3+\hat{t}^2\hat{u}
+\hat{t}\hat{u}^2+\hat{u}^3)
\Bigr)
\over (\hat{s}-m_J^2)^2(\hat{t}-m_J^2)^2(\hat{u}-m_J^2)^2},
\eeq
\beq
\hat{\sigma}_{ND1}&=&-{1\over \hat{s}\hat{u}}
{256m_J^2\over (\hat{s}-m_J^2)^3(\hat{t}-m_J^2)^3(\hat{u}-m_J^2)^3}
\nonumber\\
&&\times\Bigl(-4\hat{t}^7-32\hat{t}^6\hat{u}-89\hat{t}^5\hat{u}^2-142\hat{t}^4\hat{u}^3
-135\hat{t}^3\hat{u}^4-76\hat{t}^2\hat{u}^5-22\hat{t}\hat{u}^6-4\hat{u}^7
\nonumber\\
&&+m_J^2(20\hat{t}^6+119\hat{t}^5\hat{u}+268\hat{t}^4\hat{u}^2+330\hat{t}^3\hat{u}^3
+239\hat{t}^2\hat{u}^4+96\hat{t}\hat{u}^5+22\hat{u}^6) 
\nonumber\\
&&-m_J^4(38\hat{t}^5+168\hat{t}^4\hat{u}+291\hat{t}^3\hat{u}^2+264\hat{t}^2\hat{u}^3
+135\hat{t}\hat{u}^4+40\hat{u}^5) 
\nonumber\\
&&+m_J^6(34\hat{t}^4+106\hat{t}^3\hat{u}+127\hat{t}^2\hat{u}^2+78\hat{t}\hat{u}^3
+31\hat{u}^4) 
\nonumber\\
&&-2m_J^8(7\hat{t}^3+12\hat{t}^2\hat{u}+8\hat{t}\hat{u}^2+5\hat{u}^3)
+m_J^{10}(2\hat{t}^2-\hat{t}\hat{u}+\hat{u}^2)
\Bigr),
\eeq
\beq
\hat{\sigma}_{ND2}&=&{1\over \hat{s}\hat{u}}
{256m_J^2\over (\hat{s}-m_J^2)^3(\hat{t}-m_J^2)^3(\hat{u}-m_J^2)^3}
\nonumber\\
&&\times\Bigl(-4\hat{t}^7-32\hat{t}^6\hat{u}-91\hat{t}^5\hat{u}^2-146\hat{t}^4\hat{u}^3
-125\hat{t}^3\hat{u}^4-44\hat{t}^2\hat{u}^5+6\hat{t}\hat{u}^6+4\hat{u}^7
\nonumber\\
&&+m_J^2(20\hat{t}^6+121\hat{t}^5\hat{u}+264\hat{t}^4\hat{u}^2+286\hat{t}^3\hat{u}^3
+141\hat{t}^2\hat{u}^4+8\hat{t}\hat{u}^5-6\hat{u}^6) 
\nonumber\\
&&-m_J^4(42\hat{t}^5+184\hat{t}^4\hat{u}+277\hat{t}^3\hat{u}^2+168\hat{t}^2\hat{u}^3
+21\hat{t}\hat{u}^4) 
\nonumber\\
&&+m_J^6(46\hat{t}^4+134\hat{t}^3\hat{u}+113\hat{t}^2\hat{u}^2+18\hat{t}\hat{u}^3
+5\hat{u}^4) 
\nonumber\\
&&-2m_J^8(13\hat{t}^3+20\hat{t}^2\hat{u}+6\hat{t}\hat{u}^2+3\hat{u}^3)
+m_J^{10}(6\hat{t}^2+\hat{t}\hat{u}+3\hat{u}^2)
\Bigr),
\eeq
\beq
\hat{\sigma}_{H1}&=&{1\over \hat{s}\hat{u}}
{256m_J^2\over (\hat{s}-m_J^2)^3(\hat{t}-m_J^2)^3(\hat{u}-m_J^2)^3}
\nonumber\\
&&\times\Bigl(-2\hat{t}^7-12\hat{t}^6\hat{u}-33\hat{t}^5\hat{u}^2-56\hat{t}^4\hat{u}^3
-61\hat{t}^3\hat{u}^4-46\hat{t}^2\hat{u}^5-22\hat{t}\hat{u}^6-8\hat{u}^7
+m_J^2(10\hat{t}^6+47\hat{t}^5\hat{u}
\nonumber\\
&&+98\hat{t}^4\hat{u}^2+122\hat{t}^3\hat{u}^3 
+99\hat{t}^2\hat{u}^4+60\hat{t}\hat{u}^5+26\hat{u}^6)
-m_J^4(20\hat{t}^5+68\hat{t}^4\hat{u}+97\hat{t}^3\hat{u}^2+84\hat{t}^2\hat{u}^3
+61\hat{t}\hat{u}^4
\nonumber\\
&&+34\hat{u}^5) 
+m_J^6(20\hat{t}^4+42\hat{t}^3\hat{u}+33\hat{t}^2\hat{u}^2+26\hat{t}\hat{u}^3+23\hat{u}^4) 
-2m_J^8(5\hat{t}^3+4\hat{t}^2\hat{u}+\hat{t}\hat{u}^2+4\hat{u}^3) 
\nonumber\\
&&+m_J^{10}(2\hat{t}^2-\hat{t}\hat{u}+\hat{u}^2) 
\Bigr),
\eeq
\beq
\hat{\sigma}_{H2}&=&-{1\over \hat{s}\hat{u}}
{256m_J^2\over (\hat{s}-m_J^2)^3(\hat{t}-m_J^2)^3(\hat{u}-m_J^2)^3}
\nonumber\\
&&\times\Bigl(-4\hat{t}^7-24\hat{t}^6\hat{u}-67\hat{t}^5\hat{u}^2-112\hat{t}^4\hat{u}^3 
-119\hat{t}^3\hat{u}^4-82\hat{t}^2\hat{u}^5-34\hat{t}\hat{u}^6-8\hat{u}^7
+m_J^2(20\hat{t}^6+99\hat{t}^5\hat{u}
\nonumber\\
&&+220\hat{t}^4\hat{u}^2+282\hat{t}^3\hat{u}^3
+221\hat{t}^2\hat{u}^4+108\hat{t}\hat{u}^5+30\hat{u}^6) 
-m_J^4(40\hat{t}^5+156\hat{t}^4\hat{u}+261\hat{t}^3\hat{u}^2+244\hat{t}^2\hat{u}^3
\nonumber\\
&&+139\hat{t}\hat{u}^4+46\hat{u}^5) 
+m_J^6(40\hat{t}^4+114\hat{t}^3\hat{u}+133\hat{t}^2\hat{u}^2+90\hat{t}\hat{u}^3+37\hat{u}^4) 
-4m_J^8(5\hat{t}^3+9\hat{t}^2\hat{u}+7\hat{t}\hat{u}^2
\nonumber\\
&&+4\hat{u}^3) 
+m_J^{10}(4\hat{t}^2+3\hat{t}\hat{u}+3\hat{u}^2)
\Bigr),
\eeq
\beq
\hat{\sigma}_{H3}&=&-{\hat{t}\over \hat{s}\hat{u}}
{512m_J^2\over (\hat{s}-m_J^2)^3(\hat{t}-m_J^2)^3(\hat{u}-m_J^2)^3}
\nonumber\\
&&\times\Bigl(-\hat{t}(\hat{t}+2\hat{u})(\hat{t}^2+2\hat{t}\hat{u}+2\hat{u}^2)^2 
+m_J^2(5\hat{t}^5+23\hat{t}^4\hat{u}+43\hat{t}^3\hat{u}^2+38\hat{t}^2\hat{u}^3
+10\hat{t}\hat{u}^4-4\hat{u}^5) 
\nonumber\\
&&-m_J^4(10\hat{t}^4+32\hat{t}^3\hat{u}+33\hat{t}^2\hat{u}^2+4\hat{t}\hat{u}^3-10\hat{u}^4) 
+m_J^6(10\hat{t}^3+18\hat{t}^2\hat{u}+\hat{t}\hat{u}^2-10\hat{u}^3) 
\nonumber\\
&&+m_J^8(-5\hat{t}^2-2\hat{t}\hat{u}+5\hat{u}^2)+m_J^{10}(\hat{t}-\hat{u})
\Bigr).
\eeq


\section{Numerical simulations of the SSA in the $J/\psi$ production}

We perform numerical simulations of the $J/\psi$ $A_N$,
\beq
A_N={d\Delta\sigma\over d^3P_{J}}\Bigr/{d\sigma\over d^3P_{J}}.
\eeq
at the RHIC energy, $\sqrt{S}=200\ {\rm GeV}$\cite{PHENIX:2010hqq},
and the LHCspin energy, $\sqrt{S}=115\ {\rm GeV}$\cite{Aidala:2019pit}.
The $C$-even function $N(x_1,x_2)$ has not yet been well constrained by experiments.
Our simulations aim to clarify which hard scattering cross sections can be enhanced in these experiments
 by using the same model for the five types of the functions in (\ref{polarized}),
\beq
N(x,x)=N(x,0)=N(x,Hx)=N(x-Hx,x)=N(Hx,Hx-x)=0.002xG(x).
\eeq
We use GJR08 LO\cite{Gluck:2007ck} for the unpolarized gluon distribution function $G(x)$.
This model function reproduces the upper bound of existing experimental results 
\cite{Liu:2009zzw,PHENIX:2021irw,PHENIX:2022znm}.
\vspace{-10mm}
\begin{figure}[H]
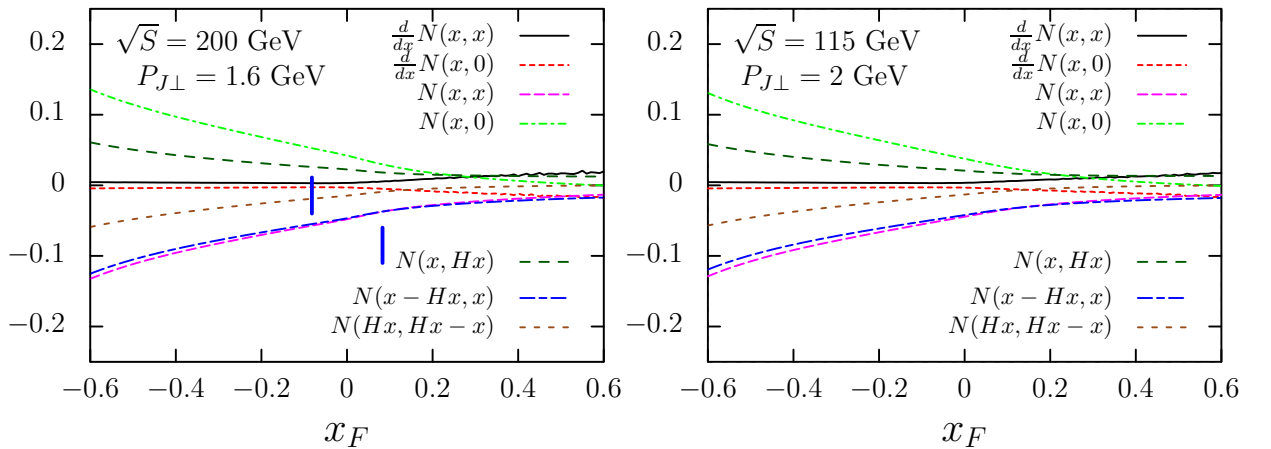

\begin{center}
\begin{minipage}{0.45\hsize}
  \input{Jpsi_SSA_xf_RHIC.tex}
\end{minipage}
\begin{minipage}{0.45\hsize}
  \input{Jpsi_SSA_xf_LHC.tex}
\end{minipage}
\end{center}
\caption{The left(right) figure shows the result of the $J/\psi$ $A_N$ for the RHIC(LHC) energy.
  The contribution from each term in (\ref{polarized}) is separately plotted. The thick error bars 
  in the left figure are taken from \cite{PHENIX:2010hqq}.}
\end{figure}

\noindent
Our calculation shows that the $J/\psi$ $A_N$ does not exhibit the typical behavior observed 
in light hadron production, where $A_N$ increases as $x_F$ becomes larger in the positive region.
This behavior can be reproduced by the derivative term of the twist-3 distribution function in light 
hadron production, and a similar behavior has also been observed in $D$-meson production. 
However, in the present case, the derivative term is found to be small over the entire range of $x_F$.
No significant difference from the RHIC results is observed in the kinematics expected for LHCspin.
\vspace{-10mm}
\begin{figure}[H]
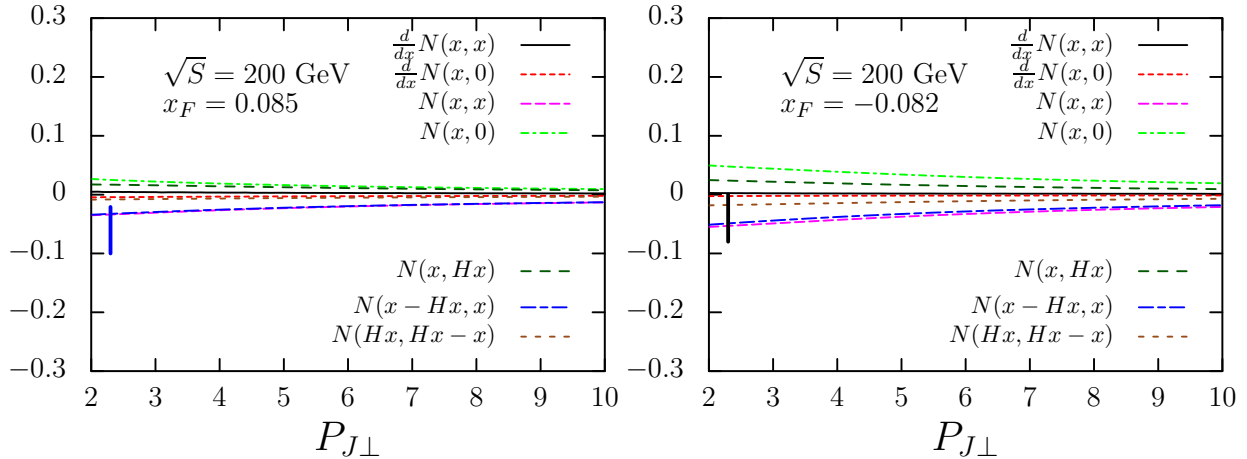

\begin{center}
\begin{minipage}{0.45\hsize}
  \input{Jpsi_SSA_pt_RHIC_pxF.tex}
\end{minipage}
\begin{minipage}{0.45\hsize}
  \input{Jpsi_SSA_pt_RHIC_nxF.tex}
\end{minipage}
\end{center}
\caption{The left(right) figure shows the $P_{J\perp}$-dependence of the $J/\psi$ $A_N$
in positive(negative) $x_F$ regions. The thick error bars are taken from \cite{PHENIX:2010hqq}.}
\end{figure}
\noindent
RHIC experiment measured the $P_{J\perp}$-dependence of the $J/\psi$ $A_N$ and
observed a nonzero value in positive $x_F$ region. LHCspin experiment is expected to cover a wide range of 
$P_{J\perp}$, $P_{J\perp}<18$ GeV for charmonium production. 
Our result will give a basis to further analyses in future experiments and lead to a phenomenological
understanding of the poorly known gluon Sivers effect.


\section{Summary}

We have derived the first result of the twist-3 gluonic contribution to the $J/\psi$ SSA in $pp$ collisions
in the color-singlet case. We find that the contribution from the $C$-odd type function 
$O(x_1,x_2)$ is completely canceled in the polarized cross section similar to the case of $ep$ 
collision and, therefore, this is another ideal observable 
to pin down the $C$-even type function $N(x_1,x_2)$.
The derivative terms ${d\over dx}N(x,x)$ and ${d\over dx}N(x,0)$, which are dominant sources
of the SSA in $D$-meson production, also appear in $J/\psi$ production.
We have performed numerical simulations at RHIC and LHC energies 
using the same model as that employed in $D$-meson production. 
Our results suggest that the contribution from the derivative terms
is small even in the large $x_F$ regions, which differs from the typical behavior 
observed in light hadron and $D$-meson production.
The RHIC experiment confirmed nonzero $A_N$, suggesting non-negligible orbital motion of gluons 
inside the proton. The LHCspin experiment has the potential to expand the accessible kinematic 
region and is essential for a more complete understanding of the poorly known gluon Sivers function.
Our results provide a basis for future investigations of the twist-3 gluon distribution function
which is responsible for describing the $J/\psi$ SSA in high-$P_{J\perp}$ regions.


\appendix

\section{Explicit forms of the hard parts}

Here we show the explicit forms of ${\cal M}^{(\pm)[\mp],ac}_{L\mu\lambda}$, which are needed for the calculation of the polarized cross section (\ref{polarized}).
We assign the color and Lorentz indices in each diagram as follows:
\begin{figure}[H]
\begin{center}
  \includegraphics[height=5.5cm,width=5.5cm]{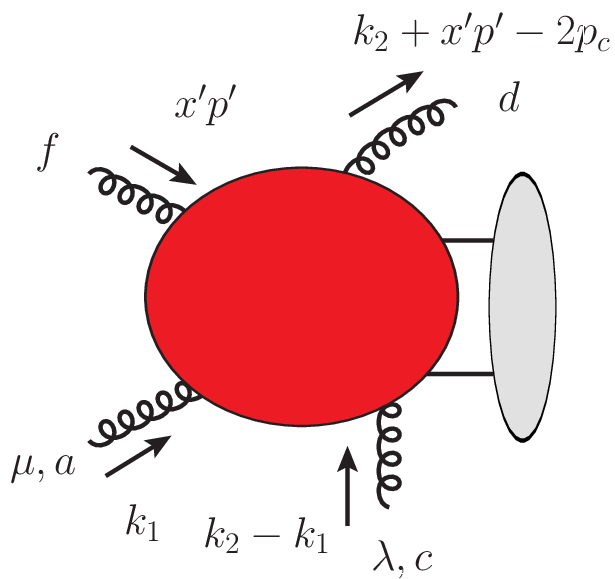}
\end{center}
 \caption{Labels of external gluon lines in ${\cal M}_{L\mu\lambda}^{ac}(k_1,k_2-k_1)$.}
\label{gluon_label}
\end{figure}
There are 86 diagrams ${\cal M}^{ac}_{Li\,\mu\lambda}\ (i=1,2,\cdots,86)$,
where the color indices $d$ and $f$ are suppressed for simplicity.
\begin{figure}[H]
\begin{center}
  \includegraphics[height=3.5cm,width=13cm]{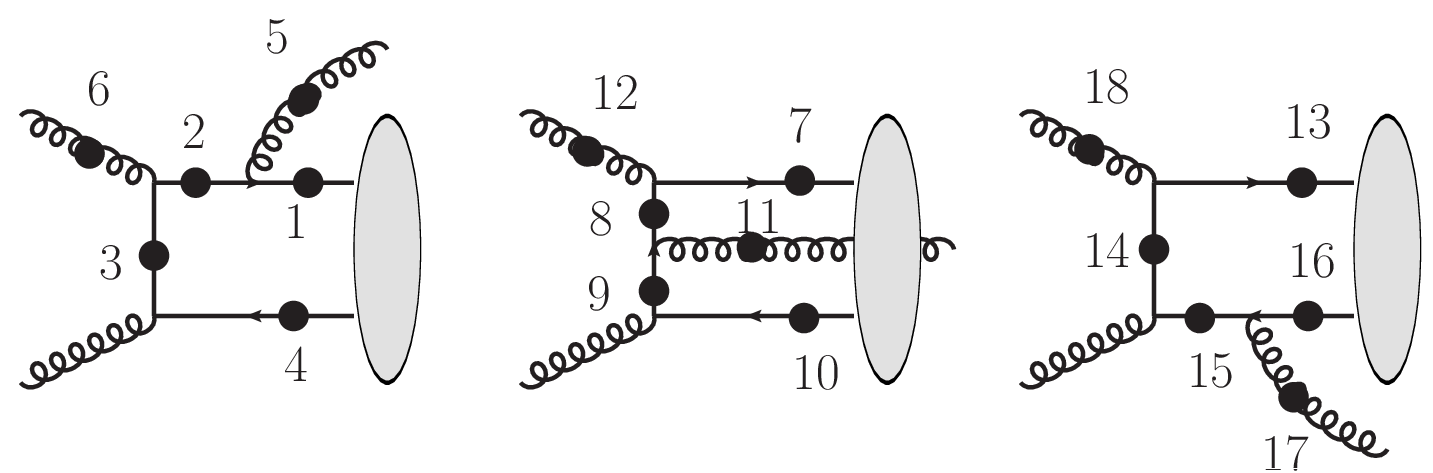}
  \includegraphics[height=3.5cm,width=9.5cm]{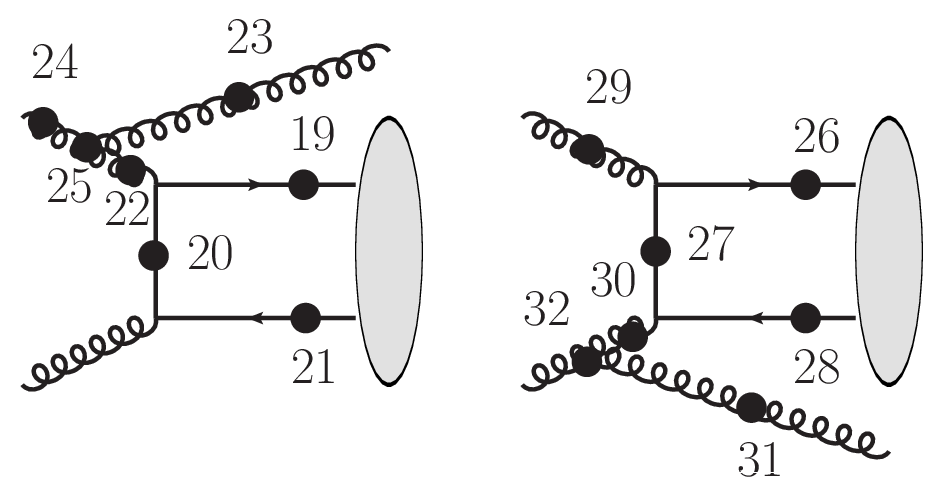}
  \includegraphics[height=3.5cm,width=13cm]{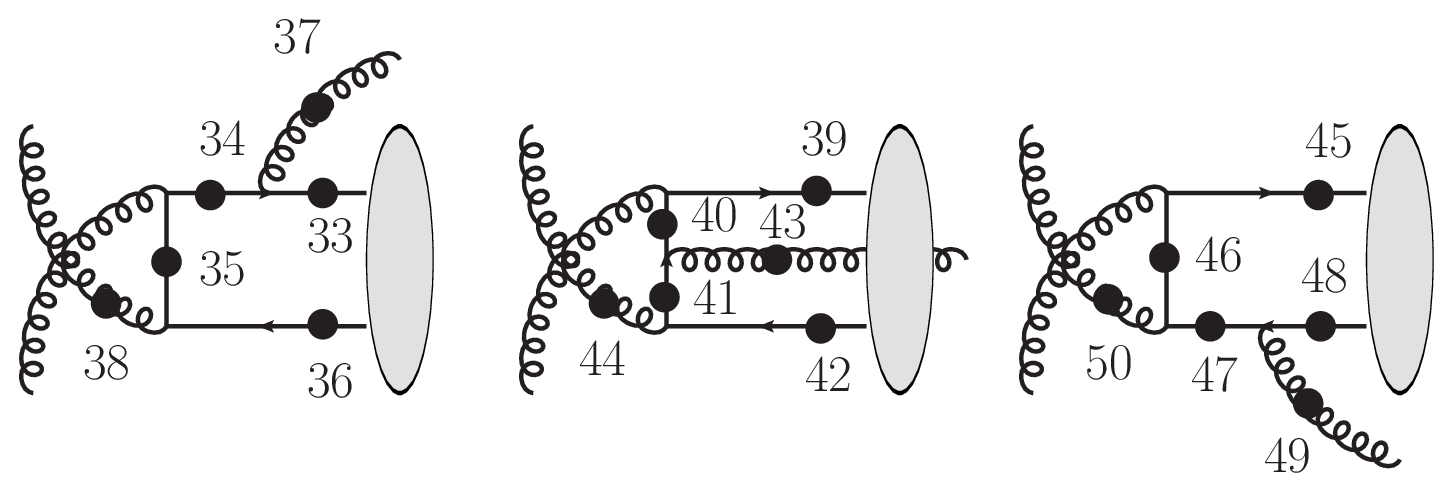}
  \includegraphics[height=3.5cm,width=9.5cm]{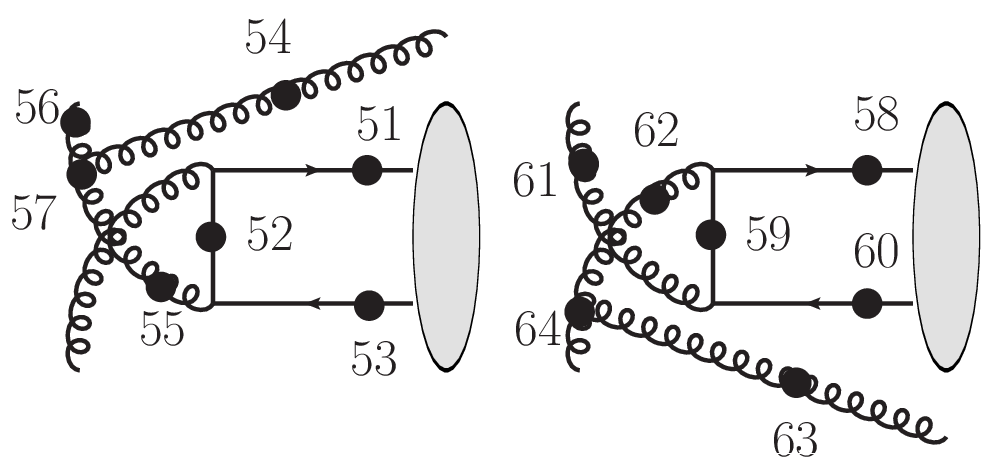}
\end{center}
\end{figure}
\begin{figure}[H]
\begin{center}
  \includegraphics[height=3.5cm,width=9.5cm]{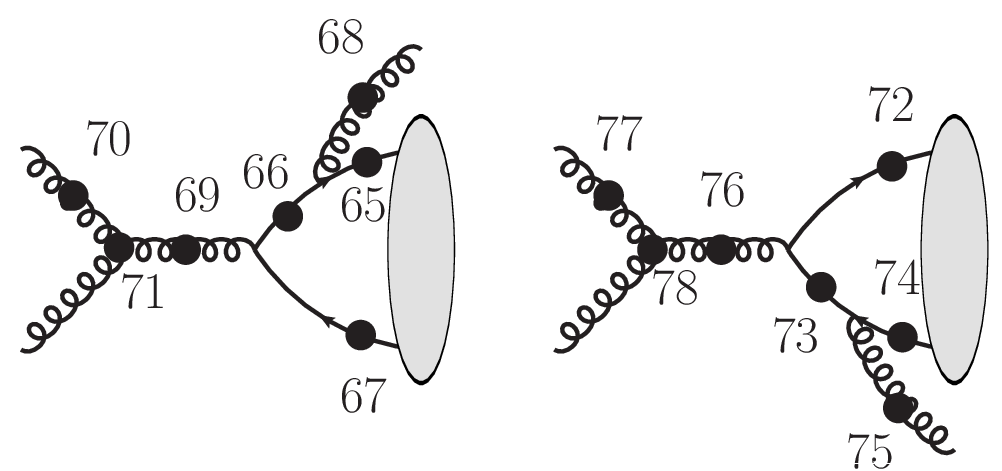}
  \includegraphics[height=3.5cm,width=9.5cm]{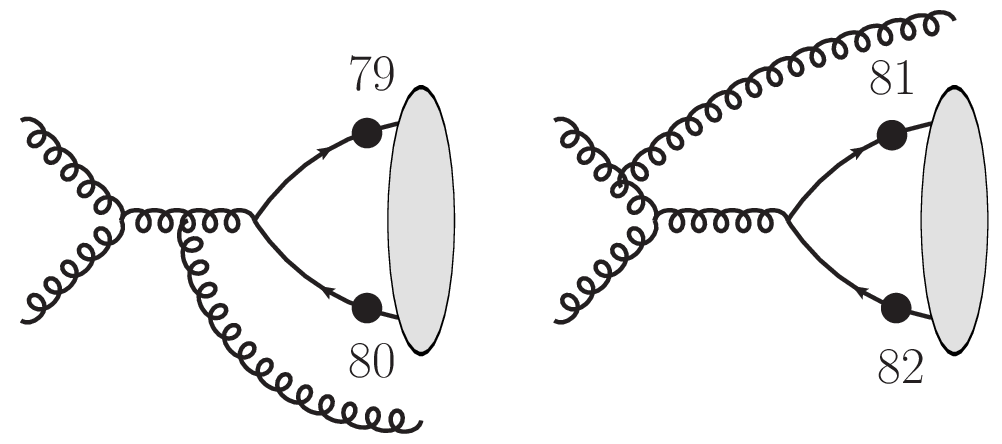}
  \includegraphics[height=3.5cm,width=9.5cm]{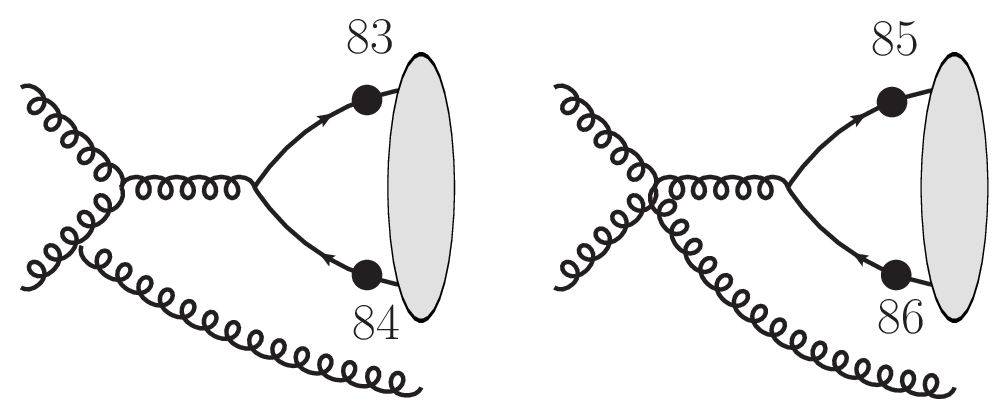}
\end{center}
 \caption{The gluon with momentum $k_2-k_1$ is connected to one of the black dots.
 The diagram with the $k_2-k_1$ gluon line connected to the $i$-th black dot gives 
 ${\cal M}_{Li\,\mu\lambda}^{ac}(k_1,k_2-k_1)$.}
\label{dynamical_dots}
\end{figure}
The diagrams including a 4-gluon vertex can be decomposed as follows:
\beq
{\cal M}_{L25\,\mu\lambda}^{ac}&=&(-if^{ced})(-if^{feg}){\rm Tr}[T^gT^a]{\cal M}^1_{25\mu\lambda}
+(-if^{cfe})(-if^{edg}){\rm Tr}[T^gT^a]{\cal M}^2_{25\mu\lambda},
\nonumber\\
{\cal M}_{L32\,\mu\lambda}^{ac}&=&(-if^{ced})(-if^{age}){\rm Tr}[T^fT^g]{\cal M}^1_{32\mu\lambda}
+(-if^{cfe})(-if^{agd}){\rm Tr}[T^eT^g]{\cal M}^2_{32\mu\lambda},
\nonumber\\
{\cal M}_{L57\,\mu\lambda}^{ac}&=&(-if^{feg})(-if^{ced}){\rm Tr}[T^aT^g]{\cal M}^1_{57\mu\lambda}
+(-if^{cfe})(-if^{edg}){\rm Tr}[T^aT^g]{\cal M}^2_{57\mu\lambda},
\nonumber\\
{\cal M}_{L64\,\mu\lambda}^{ac}&=&(-if^{age})(-if^{ced}){\rm Tr}[T^gT^f]{\cal M}^1_{64\mu\lambda}
+(-if^{cfe})(-if^{agd}){\rm Tr}[T^eT^g]{\cal M}^2_{64\mu\lambda},
\nonumber\\
{\cal M}_{L71\,\mu\lambda}^{ac}&=&(-if^{afe})(-if^{cgd}){\rm Tr}[T^gT^e]{\cal M}^1_{71\mu\lambda}
+(-if^{cfe})(-if^{aeg}){\rm Tr}[T^dT^g]{\cal M}^2_{71\mu\lambda},
\nonumber\\
{\cal M}_{L78\,\mu\lambda}^{ac}&=&(-if^{afe})(-if^{cgd}){\rm Tr}[T^gT^e]{\cal M}^1_{78\mu\lambda}
+(-if^{cfe})(-if^{aeg}){\rm Tr}[T^gT^d]{\cal M}^2_{78\mu\lambda},
\nonumber\\
{\cal M}_{L85\,\mu\lambda}^{ac}&=&(-if^{afe})(-if^{deg}){\rm Tr}[T^cT^g]{\cal M}^1_{85\mu\lambda}
+(-if^{aeg})(-if^{fde}){\rm Tr}[T^cT^g]{\cal M}^2_{85\mu\lambda},
\nonumber\\
{\cal M}_{L86\,\mu\lambda}^{ac}&=&(-if^{afe})(-if^{deg}){\rm Tr}[T^cT^g]{\cal M}^1_{86\mu\lambda}
+(-if^{aeg})(-if^{fde}){\rm Tr}[T^cT^g]{\cal M}^2_{86\mu\lambda},
\eeq
where $T^a$ is a generator of the SU($N_c$) group.
${\cal M}^{(\pm)[\mp]\,ac}_{L\mu\lambda}$ in the polarized cross section (\ref{polarized})
are given by
\beq
{\cal M}^{(+)[+]\,ac}_{L\mu\lambda}(k_1,k_2-k_1)=0,
\eeq
\beq
&&{\cal M}^{(-)[-]\,ac}_{L\mu\lambda}(k_1,k_2-k_1)
\nonumber\\
&=&{\rm Tr}[T^cT^dT^fT^a]{\cal M}_{1\mu\lambda}
+{\rm Tr}[T^dT^cT^fT^a]{\cal M}_{2\mu\lambda}
+{\rm Tr}[T^aT^dT^cT^f]{\cal M}_{3\mu\lambda}
+{\rm Tr}[T^dT^aT^fT^c]{\cal M}_{4\mu\lambda}
\nonumber\\
&&+(-if^{ced}){\rm Tr}[T^aT^eT^f]{\cal M}_{5\mu\lambda}
+{\rm Tr}[T^cT^fT^dT^a]{\cal M}_{7\mu\lambda}
+{\rm Tr}[T^cT^fT^dT^a]{\cal M}_{8\mu\lambda}
\nonumber\\
&&+\Bigl({\rm Tr}[T^cT^fT^dT^a]-(-if^{ced}){\rm Tr}[T^fT^eT^a]\Bigr){\cal M}_{9\mu\lambda}
+\Bigl({\rm Tr}[T^aT^fT^dT^c]-(-if^{aed}){\rm Tr}[T^fT^eT^c]\Bigr){\cal M}_{10\mu\lambda}
\nonumber\\
&&+(-if^{ced}){\rm Tr}[T^aT^fT^e]\Bigl({\cal M}_{11\mu\lambda}
+{\rm Tr}[T^cT^fT^aT^d]{\cal M}_{13\mu\lambda}
+{\rm Tr}[T^cT^fT^aT^d]{\cal M}_{14\mu\lambda}
\nonumber\\
&&+{\rm Tr}[T^aT^fT^cT^d]{\cal M}_{15\mu\lambda}
+{\rm Tr}[T^aT^fT^dT^c]{\cal M}_{16\mu\lambda}
+(-if^{ced}){\rm Tr}[T^aT^fT^e]{\cal M}_{17\mu\lambda}
\nonumber\\
&&+(-if^{fde}){\rm Tr}[T^cT^eT^a]{\cal M}_{19\mu\lambda}
+\Bigl((-if^{fde}){\rm Tr}[T^eT^cT^a]-(-if^{cfe})(-if^{edg}){\rm Tr}[T^gT^a]\Bigr)
{\cal M}_{20\mu\lambda}
\nonumber\\
&&+\Bigl((-if^{fde}){\rm Tr}[T^aT^eT^c]+(-if^{aed})(-if^{fge}){\rm Tr}[T^gT^c]
\Bigr){\cal M}_{21\mu\lambda}
-(-if^{ced})(-if^{feg}){\rm Tr}[T^gT^a]{\cal M}_{22\mu\lambda}
\nonumber\\
&&+(-if^{ced})(-if^{feg}){\rm Tr}[T^gT^a]{\cal M}_{23\mu\lambda}
+(-if^{ced})(-if^{feg}){\rm Tr}[T^gT^a]{\cal M}^1_{25\mu\lambda}
+(-if^{aed}){\rm Tr}[T^cT^fT^e]{\cal M}_{26\mu\lambda}
\nonumber\\
&&+(-if^{aed}){\rm Tr}[T^cT^fT^e]{\cal M}_{27\mu\lambda}
+(-if^{aed}){\rm Tr}[T^fT^eT^c]{\cal M}_{28\mu\lambda}
+{\rm Tr}[T^cT^dT^fT^a]{\cal M}_{33\mu\lambda}
\nonumber\\
&&+{\rm Tr}[T^dT^cT^fT^a]{\cal M}_{34\mu\lambda}
+{\rm Tr}[T^cT^dT^aT^f]{\cal M}_{35\mu\lambda}
+{\rm Tr}[T^dT^aT^fT^c]{\cal M}_{36\mu\lambda}
+(-if^{ced}){\rm Tr}[T^aT^eT^f]{\cal M}_{37\mu\lambda}
\nonumber\\
&&+\Bigr({\rm Tr}[T^cT^dT^fT^a]+(-if^{aed}){\rm Tr}[T^cT^eT^f]
\Bigr){\cal M}_{39\mu\lambda}
+\Bigl({\rm Tr}[T^aT^cT^dT^f]+(-if^{cfe}){\rm Tr}[T^aT^dT^e]\Bigr)
{\cal M}_{40\mu\lambda}
\nonumber\\
&&+{\rm Tr}[T^cT^aT^dT^f]{\cal M}_{41\mu\lambda}
+{\rm Tr}[T^aT^dT^fT^c]{\cal M}_{42\mu\lambda}
+(-if^{ced}){\rm Tr}[T^aT^eT^f]{\cal M}_{43\mu\lambda}
+{\rm Tr}[T^cT^fT^aT^d]{\cal M}_{45\mu\lambda}
\nonumber\\
&&+{\rm Tr}[T^aT^fT^cT^d]{\cal M}_{46\mu\lambda}
+{\rm Tr}[T^aT^fT^cT^d]{\cal M}_{47\mu\lambda}
+{\rm Tr}[T^aT^fT^dT^c]{\cal M}_{48\mu\lambda}
+(-if^{ced}){\rm Tr}[T^aT^fT^e]{\cal M}_{49\mu\lambda}
\nonumber\\
&&+\Bigl((-if^{fde}){\rm Tr}[T^cT^eT^a]-(-if^{aed})(-if^{fge}){\rm Tr}[T^cT^g]
\Bigr){\cal M}_{51\mu\lambda}
\nonumber\\
&&+\Bigl((-if^{fde}){\rm Tr}[T^cT^aT^e]+(-if^{feg})(-if^{ced}){\rm Tr}[T^aT^g]
\Bigr){\cal M}_{52\mu\lambda}
+(-if^{fde}){\rm Tr}[T^aT^eT^c]{\cal M}_{53\mu\lambda}
\nonumber\\
&&+(-if^{ced})(-if^{feg}){\rm Tr}[T^gT^a]{\cal M}_{54\mu\lambda}
-(-if^{ced})(-if^{feg}){\rm Tr}[T^gT^a]{\cal M}_{55\mu\lambda}
\nonumber\\
&&+(-if^{ced})(-if^{feg}){\rm Tr}[T^gT^a]{\cal M}^1_{57\mu\lambda}
+(-if^{aed}){\rm Tr}[T^cT^eT^f]{\cal M}_{58\mu\lambda}
+(-if^{aed}){\rm Tr}[T^cT^eT^f]{\cal M}_{59\mu\lambda}
\nonumber\\
&&+(-if^{aed}){\rm Tr}[T^eT^fT^c]{\cal M}_{60\mu\lambda}
+(-if^{aed})(-if^{fge}){\rm Tr}[T^cT^g]{\cal M}_{81\mu\lambda}
+(-if^{aed})(-if^{fge}){\rm Tr}[T^gT^c]{\cal M}_{82\mu\lambda}
\nonumber\\
&&+(-if^{aed})(-if^{fge}){\rm Tr}[T^cT^g]{\cal M}_{83\mu\lambda}
+(-if^{aed})(-if^{fge}){\rm Tr}[T^gT^c]{\cal M}_{84\mu\lambda}
\nonumber\\
&&+(-if^{aed})(-if^{fge}){\rm Tr}[T^gT^c]{\cal M}^1_{85\mu\lambda}
+(-if^{aed})(-if^{fge}){\rm Tr}[T^cT^g]{\cal M}^2_{85\mu\lambda},
\eeq
\beq
&&{\cal M}^{(-)[+]\,ac}_{L\mu\lambda}(k_1,k_2-k_1)
\nonumber\\
&=&(-if^{cfe})\Bigl[{\rm Tr}[T^dT^eT^a]{\cal M}_{3\mu\lambda}
+{\rm Tr}[T^dT^eT^a]{\cal M}_{6\mu\lambda}
+{\rm Tr}[T^eT^dT^a]{\cal M}_{8\mu\lambda}
+{\rm Tr}[T^eT^dT^a]{\cal M}_{9\mu\lambda}
\nonumber\\
&&+{\rm Tr}[T^aT^eT^d]{\cal M}_{12\mu\lambda}
+{\rm Tr}[T^eT^aT^d]{\cal M}_{14\mu\lambda}
+{\rm Tr}[T^aT^dT^e]{\cal M}_{18\mu\lambda}
+(-if^{edg}){\rm Tr}[T^gT^a]{\cal M}_{20\mu\lambda}
\nonumber\\
&&+(-if^{edg}){\rm Tr}[T^aT^g]{\cal M}_{22\mu\lambda}
+(-if^{edg}){\rm Tr}[T^aT^g]{\cal M}_{24\mu\lambda}
+(-if^{edg}){\rm Tr}[T^gT^a]{\cal M}^2_{25\mu\lambda}
\nonumber\\
&&+(-if^{agd}){\rm Tr}[T^eT^g]{\cal M}_{27\mu\lambda}
+(-if^{agd}){\rm Tr}[T^eT^g]{\cal M}_{29\mu\lambda}
-(-if^{agd}){\rm Tr}[T^eT^g]{\cal M}_{30\mu\lambda}
\nonumber\\
&&+(-if^{agd}){\rm Tr}[T^eT^g]{\cal M}^2_{32\mu\lambda}
-{\rm Tr}[T^dT^aT^e]{\cal M}_{35\mu\lambda}
+{\rm Tr}[T^dT^aT^e]({\cal M}_{38\mu\lambda}
-{\rm Tr}[T^aT^dT^e]{\cal M}_{40\mu\lambda}
\nonumber\\
&&-{\rm Tr}[T^aT^dT^e]{\cal M}_{41\mu\lambda}
+{\rm Tr}[T^aT^dT^e]{\cal M}_{44\mu\lambda}
-{\rm Tr}[T^aT^eT^d]{\cal M}_{46\mu\lambda}
+{\rm Tr}[T^aT^eT^d]{\cal M}_{50\mu\lambda}
\nonumber\\
&&-(-if^{edg}){\rm Tr}[T^aT^g]{\cal M}_{52\mu\lambda}
+(-if^{edg}){\rm Tr}[T^aT^g]{\cal M}_{55\mu\lambda}
+(-if^{edg}){\rm Tr}[T^aT^g]{\cal M}_{56\mu\lambda}
\nonumber\\
&&+(-if^{edg}){\rm Tr}[T^gT^a]{\cal M}^2_{57\mu\lambda}
+(-if^{adg}){\rm Tr}[T^gT^e]{\cal M}_{59\mu\lambda}
-(-if^{adg}){\rm Tr}[T^gT^e]{\cal M}_{61\mu\lambda}
\nonumber\\
&&+(-if^{adg}){\rm Tr}[T^eT^g]{\cal M}_{62\mu\lambda}
+(-if^{adg}){\rm Tr}[T^eT^g]{\cal M}^2_{64\mu\lambda}
+(-if^{aeg}){\rm Tr}[T^dT^g]{\cal M}_{70\mu\lambda}
\nonumber\\
&&-(-if^{aeg}){\rm Tr}[T^dT^g]{\cal M}^2_{71\mu\lambda}
-(-if^{adg}){\rm Tr}[T^eT^g]{\cal M}_{77\mu\lambda}
-(-if^{adg}){\rm Tr}[T^eT^g]{\cal M}^2_{78\mu\lambda}
\Bigr],
\eeq
\beq
&&{\cal M}^{(+)[-]\,ac}_{L\mu\lambda}(k_1,k_2-k_1)
\nonumber\\
&=&(-if^{afe})\Bigl[{\rm Tr}[T^dT^eT^c]{\cal M}_{4\mu\lambda}
+{\rm Tr}[T^eT^dT^c]{\cal M}_{10\mu\lambda}
+{\rm Tr}[T^eT^cT^d]{\cal M}_{15\mu\lambda}
+{\rm Tr}[T^eT^dT^c]{\cal M}_{16\mu\lambda}
\nonumber\\
&&+(-if^{cgd}){\rm Tr}[T^eT^g]{\cal M}_{17\mu\lambda}
-(-if^{deg}){\rm Tr}[T^gT^c]{\cal M}_{21\mu\lambda}
+(-if^{cgd}){\rm Tr}[T^eT^g]{\cal M}_{31\mu\lambda}
\nonumber\\
&&
+(-if^{cgd}){\rm Tr}[T^eT^g]{\cal M}^1_{32\mu\lambda}
-{\rm Tr}[T^cT^dT^e]{\cal M}_{33\mu\lambda}
-{\rm Tr}[T^dT^cT^e]{\cal M}_{34\mu\lambda}
-(-if^{cgd}){\rm Tr}[T^gT^e]{\cal M}_{37\mu\lambda}
\nonumber\\
&&
-{\rm Tr}[T^cT^dT^e]{\cal M}_{39\mu\lambda}
-{\rm Tr}[T^cT^eT^d]{\cal M}_{45\mu\lambda}
+(-if^{deg}){\rm Tr}[T^cT^g]{\cal M}_{51\mu\lambda}
+(-if^{cgd}){\rm Tr}[T^gT^e]{\cal M}_{63\mu\lambda}
\nonumber\\
&&
+(-if^{cgd}){\rm Tr}[T^gT^e]{\cal M}^1_{64\mu\lambda}
+{\rm Tr}[T^cT^dT^e]{\cal M}_{65\mu\lambda}
+{\rm Tr}[T^dT^cT^e]{\cal M}_{66\mu\lambda}
+{\rm Tr}[T^dT^eT^c]{\cal M}_{67\mu\lambda}
\nonumber\\
&&
+(-if^{cgd}){\rm Tr}[T^gT^e]{\cal M}_{68\mu\lambda}
+(-if^{ceg}){\rm Tr}[T^dT^g]{\cal M}_{69\mu\lambda}
+(-if^{ceg}){\rm Tr}[T^dT^g]{\cal M}^1_{71\mu\lambda}
\nonumber\\
&&+{\rm Tr}[T^cT^eT^d]{\cal M}_{72\mu\lambda}
+{\rm Tr}[T^eT^cT^d]{\cal M}_{73\mu\lambda}
+{\rm Tr}[T^eT^dT^c]{\cal M}_{74\mu\lambda}
+(-if^{cgd}){\rm Tr}[T^eT^g]{\cal M}_{75\mu\lambda}
\nonumber\\
&&
+(-if^{ceg}){\rm Tr}[T^gT^d]{\cal M}_{76\mu\lambda}
+(-if^{ceg}){\rm Tr}[T^gT^d]{\cal M}^1_{78\mu\lambda}
+(-if^{deg}){\rm Tr}[T^cT^g]{\cal M}_{79\mu\lambda}
\nonumber\\
&&
+(-if^{deg}){\rm Tr}[T^cT^g]{\cal M}_{80\mu\lambda}
-(-if^{deg}){\rm Tr}[T^cT^g]{\cal M}_{81\mu\lambda}
-(-if^{deg}){\rm Tr}[T^cT^g]{\cal M}_{82\mu\lambda}
\nonumber\\
&&+(-if^{deg}){\rm Tr}[T^cT^g]{\cal M}^1_{85\mu\lambda}
+(-if^{deg}){\rm Tr}[T^cT^g]{\cal M}^1_{86\mu\lambda}
\Bigr].
\eeq
They satisfy the Ward-Takahashi identity
\beq
(k_2-k_1)^{\lambda}{\cal M}^{(\pm)[\pm]\,ac}_{L\mu\lambda}(k_1,k_2-k_1)
&=&\sum C^{(\pm)[\pm]}_{Li\,ac}{\cal M}_{i\,\mu}(k_2),
\nonumber\\
k_1^{\mu}{\cal M}^{(\pm)[\pm]\,ac}_{L\mu\lambda}(k_1,k_2-k_1)
&=&-\sum C^{(\pm)[\pm]}_{Li\,ac}{\cal M}_{i\,\lambda}(k_2),
\eeq
where the color factors are given by
\beq
C^{(-)[+]}_{L1\,ac}&=&(-if^{cfe}){\rm Tr}[T^dT^eT^a],\hspace{5mm}
C^{(-)[+]}_{L2\,ac}=(-if^{cfe}){\rm Tr}[T^eT^dT^a],\hspace{5mm}
C^{(-)[+]}_{L3\,ac}=(-if^{cfe}){\rm Tr}[T^eT^aT^d],
\nonumber\\
C^{(-)[+]}_{L4\,ac}&=&(-if^{cfe})(-if^{edg}){\rm Tr}[T^gT^a],\hspace{5mm}
C^{(-)[+]}_{L5\,ac}=(-if^{cfe})(-if^{edg}){\rm Tr}[T^gT^a],\hspace{5mm}
C^{(-)[+]}_{L6\,ac}=(-if^{cfe}){\rm Tr}[T^dT^aT^e],
\nonumber\\
C^{(-)[+]}_{L7\,ac}&=&(-if^{cfe}){\rm Tr}[T^aT^dT^e],\hspace{5mm}
C^{(-)[+]}_{L8\,ac}=(-if^{cfe}){\rm Tr}[T^aT^eT^d],\hspace{5mm}
C^{(-)[+]}_{L9\,ac}=(-if^{cfe})(-if^{edg}){\rm Tr}[T^aT^g],
\nonumber\\
C^{(-)[+]}_{L10\,ac}&=&(-if^{cfe})(-if^{agd}){\rm Tr}[T^eT^g],\hspace{5mm}
C^{(-)[+]}_{L11\,ac}=C^{(-)[+]}_{L12\,ac}
=(-if^{cfe})(-if^{aeg}){\rm Tr}[T^dT^g],
\eeq
\beq
C^{(+)[-]}_{L1\,ac}&=&-(-if^{afe}){\rm Tr}[T^dT^eT^c],\hspace{5mm}
C^{(+)[-]}_{L2\,ac}=-(-if^{afe}){\rm Tr}[T^eT^dT^c],\hspace{5mm}
C^{(+)[-]}_{L3\,ac}=-(-if^{afe}){\rm Tr}[T^eT^cT^d],
\nonumber\\
C^{(+)[-]}_{L4\,ac}&=&(-if^{afe})(-if^{deg}){\rm Tr}[T^gT^c],\hspace{5mm}
C^{(+)[-]}_{L5\,ac}=-(-if^{afe})(-if^{cgd}){\rm Tr}[T^eT^g],\hspace{5mm}
\nonumber\\
C^{(+)[-]}_{L6\,ac}&=&-(-if^{afe}){\rm Tr}[T^dT^cT^e],\hspace{5mm}
C^{(+)[-]}_{L7\,ac}=-(-if^{afe}){\rm Tr}[T^cT^dT^e],
\nonumber\\
C^{(+)[-]}_{L8\,ac}&=&-(-if^{afe}){\rm Tr}[T^cT^eT^d],\hspace{5mm}
C^{(+)[-]}_{L9\,ac}=(-if^{afe})(-if^{deg}){\rm Tr}[T^cT^g],
\nonumber\\
C^{(+)[-]}_{L10\,ac}&=&-(-if^{afe})(-if^{cgd}){\rm Tr}[T^gT^e],\hspace{5mm}
C^{(+)[-]}_{L11\,ac}=C^{(+)[-]}_{L12\,ac}=-(-if^{afe})(-if^{ceg}){\rm Tr}[T^dT^g],
\eeq
\beq
C^{(-)[-]}_{L1\,ac}&=&C^{(-)[-]}_{L2\,ac}=C^{(-)[-]}_{L3\,ac}
=C^{(-)[-]}_{L6\,ac}=C^{(-)[-]}_{L7\,ac}=C^{(-)[-]}_{L8\,ac}=
-{\rm Tr}[T^dT^aT^fT^c]+{\rm Tr}[T^aT^dT^cT^f],
\nonumber\\
C^{(-)[-]}_{L4\,ac}&=&C^{(-)[-]}_{L5\,ac}=C^{(-)[-]}_{L9\,ac}
=C^{(-)[-]}_{L10\,ac}=C^{(-)[-]}_{L11\,ac}=C^{(-)[-]}_{L12\,ac}=0.
\eeq
The color factors in the polarized cross section (\ref{polarized}) are obtained as
\beq
C^{(\pm)[\pm]}_{Li\,ac}C^{b*}_j=C^{f(\pm)[\pm]}_{Lij}if^{abc}+C^{d(\pm)[\pm]}_{Lij}d^{abc}.
\eeq


\section*{Acknowledgements}

This work is supported by Polish National Science Center Grant No. UMO-2023/49/B/ST2/03665.

\end{document}